\documentclass[sigconf]{acmart}
\AtBeginDocument{%
  }

\setcopyright{acmlicensed}
\copyrightyear{2025}
\acmYear{2025}
\acmDOI{XXXXXXX.XXXXXXX}
\acmConference[ISWC]{UbiComp/ISWC}{October 14--16,
  2025}{Espoo, Finland}
\acmISBN{XXXXXXXXXXXXXXXX}

\usepackage{balance}

\begin{document}

\title{HARNode: A Time-Synchronised, Open-Source, Multi-Device, Wearable System for Ad Hoc Field Studies} 

\author{Philipp Lepold}
\email{philipp.lepold@kit.edu}
\orcid{TODO}
\affiliation{%
  \institution{Karlsruhe Institute of Technology}
  \city{Karlsruhe}
  \country{Germany}
}

\author{Tobias Röddiger}
\email{tobias.roeddiger@kit.edu}
\orcid{TODO}
\affiliation{%
  \institution{Karlsruhe Institute of Technology}
  \city{Karlsruhe}
  \country{Germany}
}

\author{Michael Beigl}
\email{michael.beigl@kit.edu}
\orcid{TODO}
\affiliation{%
  \institution{Karlsruhe Institute of Technology}
  \city{Karlsruhe}
  \country{Germany}
}

\renewcommand{\shortauthors}{Lepold et al.}

\begin{abstract}
Human activity recognition (HAR) research often lacks accessible, comprehensive field data. Commercial systems are rarely open source, hard to expand, and limited by issues like node synchronisation, data throughput, unclear sensor placement, complexity, and high cost. As a result, researchers typically use only a few intuitively placed sensors and conduct limited field trials. HARNode overcomes these challenges with a fully open-source hardware and software platform. Each node includes an ESP32-S3 module (AtomS3), a 9-axis IMU (Bosch BMX160), pressure and temperature sensors (Bosch BMP388), a display, and an I²C port. Data is streamed via Wi-Fi, with NTP-based time synchronisation achieving ~1 ms accuracy. The system runs for up to 8 hours and is built using off-the-shelf parts, a simple online PCB service, and a compact 3D-printed housing with Velcro straps, enabling flexible and scalable body placement while requiring little hardware knowledge. In a study with ten subjects wearing eleven HARNodes each, setup took under five minutes per person. A random forest classifier distinguished walking from stair-climbing transitions, showing the benefits of sensor-overprovisioning: Seven nodes achieved $\approx$ 98\% accuracy, matching the performance of all eleven. These findings confirm HARNode’s value as a fast-deploying, scalable tool for field-based HAR research and optimised sensor placement.
\end{abstract}




\maketitle

\section{Introduction and Related Work}
Wearable inertial sensors have become extremely compact and affordable, leading to their widespread use in mobile and wearable computing research. In particular, inertial measurement units (IMUs) are commonly used for human activity recognition (HAR), including classification of activities such as level walking and stair climbing \cite{pesenti2023imu, zhang2022deep}. 
A major limitation in human activity detection research and use is the availability of a large amount of collected data, especially in very specific application domains where meaningful data can only be collected in the field. In our application example, HAR using IMUs is essential for real-time gait transition detection and adaptive assistance in wearable exoskeletons to support elderly and mobility-impaired individuals \cite{kapsalyamov2019state, sposito2022exoskeletons}.

Currently, due to technical limitations - e.g. node synchronisation time and throughput - and practical considerations - setup time and cost - we mostly rely on a limited number of wearable sensors on the body, typically positioned at one or two body locations \cite{moreira2022review}. This is suboptimal, as the correct positioning of the wearable on the body is based more on intuition than evidence. 

In this paper, we present a fully open-source, extendable wearable system that addresses these issues and is tuned for fast and easy setup (including a fast setup procedure) and data collection in the field with an (almost) arbitrarily large number of highly synchronised sensor nodes. 

\begin{table}[h]
    \centering
        \caption{Comparison of adaptable wearable IMU sensor systems. Open\,= open-source; Max\,= maximum number of IMU nodes; BT\,= Bluetooth; prop.\,= proprietary; 7 nodes are a typical max. for BT}

    \resizebox{\columnwidth}{!}{%
        \begin{tabular}{|c|c|c|c|c|c|}
            \hline
            \textbf{System} & \textbf{Open} & \textbf{Max} & \textbf{Connectivity} & \textbf{Features} \\
            \hline
            OpenSenseRT \cite{opensensert_opensim}          & Yes & 6--8+  & Wired sensors          & Low cost; easy assembly \\
            \hline
            movisens EdaMove 4 \cite{movisens}  & No  & (7)    & BT, USB                & EDA; steps; waterproof \\
            \hline
            Xsens Awinda  \cite{xsens_awinda}       & No  & 17    & prop.               & High accuracy \\
            \hline
            APDM Opal \cite{apdm_opal}           & No  & 24   & prop.               & Gait analysis \\
            \hline
            Noraxon \cite{noraxon}              & No  & (7)    & propr.\,+BLE               & EMG integration \\
            \hline
            BioX Bands \cite{biox_bands_researchgate}           & No  & (7)    & BT SPP               & FSR for muscle-contraction \\
            \hline
            \textbf{HARNode} (ours)& Yes & 100+   & Wi-Fi (802.11 b/g/n)             & Fast field setup; low cost; extendable \\

            \hline
        \end{tabular}%
    }

    \label{tab:wearable_imus}
\end{table}

Existing open-source alternatives, such as OpenSenseRT \cite{slade2021open}, are constrained by wired configurations, external battery requirements, and complex setup procedures, or are commercial. \autoref{tab:wearable_imus} gives a short overview of various systems and compares them to this paper's approach. In addition to these works, specific devices have been reported, for example for mass sensing on the body \cite{tsukamoto2023best}. We have excluded such work, as it is not intended to be easily usable or reproducible by researchers.

Central to the solution presented in this paper is the use of off-the-shelf components, including an ESP32-based AtomS3 microcontroller with integrated display for status monitoring and sensor identification. 
Each node has a built-in 9-axis IMU (accelerometer, gyroscope, magnetometer), as well as barometric pressure and temperature sensors. 
All the hardware is housed in a compact 3D-printed case, which also fits the battery and is secured to the body with flexible Velcro straps, making the nodes lightweight, wireless, and easy to attach to different parts of the body. Data is streamed over Wi-Fi, and we use a high-precision NTP-based method for time synchronisation of the nodes. All elements of HARNode are completely open source, including hardware, firmware, and the server.

In addition to describing the hardware, software, system, and its use to set up field data collection, we also report a small proof-of-concept study that demonstrates the capabilities of the system. Here, we conducted a study with 10 participants, each wearing 11 HARNodes positioned on different parts of the body. The goal was to collect motion data during stair approach to systematically investigate which sensor placements are most informative for detecting transitions relevant to exoskeleton control.

\begin{figure*}[t!]
     \centering
    \includegraphics[width=\linewidth]{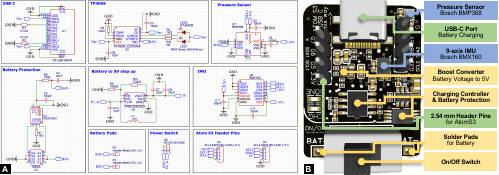}
    \caption{(A) Schematic diagram of the sensor node, showing the sensors, and power circuitry; (B) Assembled sensor node PCB with labelled components.}
    \label{fig:hardware}
\end{figure*}

\section{Challenges and Design Requirements}
Current work on human activity recognition (HAR) in the wild is constrained by data scarcity -- often addressed using additional synthetic data (e.g. \citet{hong2024crosshar}), communication and synchronisation limits \cite{contoli2024energy}, high deployment cost \cite{hussain2019different}, and ad hoc sensor placement \cite{gil2023reducing}\cite{kunze2014sensor}. Translating these observations into explicit design requirements is essential for scalable, evidence-based sensing systems. We identified the following requirements.

\begin{enumerate}
\item \textbf{Time alignment and constrained throughput}.  
Multi-node wearables drift by several milliseconds on commodity Bluetooth links, corrupting feature fusion and frequency analysis~\cite{Biagetti2025}. 

\textbf{Requirement}: Best-of-the-art commodity sub-millisecond time-stamping, high ($\gg$50kbit/s) throughput, local buffering with burst upload

\item \textbf{Deployment overhead}.  
Every additional node inflates hardware, fitting and maintenance costs; user surveys list donning time, comfort and aesthetics as primary barriers~\cite{Ni2024,Li2022}.

\textbf{Requirement}: single-board MCUs with integrated IMUs, tool-less Velcro or textile mounts, semi-automatic node configuration to cut setup below 20 seconds per node.

\item \textbf{Evidence-free sensor placement}.  
Recognition accuracy varies sharply with body location; mis-placement can halve F-scores~\cite{Davoudi2021}. 

\textbf{Requirement}: Standardised setup for systematic placement campaigns, with high number of nodes to find optimal placements, and modular housings enabling quick repositioning.

\item \textbf{Wearablity aspects}.  
Wearability aspects must be taken into account, as summarised by ~\citet{zeagler2017wear}.

\textbf{Requirement}: Hardware and housing should be appropriate for attachment, taking into account e.g. attachment points and other wearablity factors and practical factors such as battery life time.
\end{enumerate}

\section{Hard- and Software}
The HARNode Wearable system consists of three main elements: The HARNode hardware, the software (HARNode device firmware and server software), and an off-the-shelf computer or laptop as a server. The hardware and software developed in this work are open-source and tailored for ease of use, also outside the laboratory environment.

\subsection{Hardware}
We developed a PCB and platform based on the M5Stack AtomS3\footnote{\url{https://docs.m5stack.com/en/core/AtomS3}}. The PCB snaps into a custom 3D-printed housing that also fits the battery and has Velcro straps for easy attachment. Overall lifetime with a single fully charged battery is designed to be between 2-8 hours for the below given sampling rates and power consumption numbers (variations depend on battery size and Wi-Fi/RF situation). 

\subsubsection{AtomS3}
The core of the HARNode is an AtomS3 microcontroller, whose schematic and source code is also open-source under the MIT licence. It is based on the ESP32-S3 architecture (Xtensa LX7 dualcore, 300 µs/min RTC deviation \cite{espressif2021esp32s3}) and has an integrated display and button, as well as Wi-Fi (IEEE 802.11 b/g/n) and Bluetooth 5 (LE) connectivity. 
In our setup while using a sample rate of 166.67 Hz, the average power consumption is 350 mA @ 3.3 V (Wi-Fi + CPU).

\subsubsection{PCB}
An overview of the PCB and its schematic can be seen in \autoref{fig:hardware}.
The components include:
\begin{itemize}
    \item \textit{Bosch} BMX160, a 9-axis inertial measurement unit (IMU): accelerometer (up to 1600 Hz,180 µA), gyroscope (up to 6400 Hz, 850 µA), magnetometer (12 Hz, 600 µA)
    \item \textit{Bosch} BMP388 pressure sensor (up to 200 Hz, 700 µA, 2 cm resolution in height difference) and temperature sensor (200 Hz, $\pm 0.5\,^{\circ}\mathrm{C}$)
    \item \textit{NanJing Top Power ASIC} TP4056X linear lithium-ion charging controller
    \item \textit{Winsok Semiconductors} WSTDW01 battery protection circuit
    \item \textit{Texas Instruments} TPS61023 boost converter: to supply the AtomS3 with 5 V, from 3 V - 4.2 V from the battery
    \item \textit{Jauch LP523450JU or LP103450JH} 950 mAh / 5.5 mm thickness / 25 g or 1900 mAh / 10 mm / 35 g 3.7 V lithium-polymer battery (connected to the board)
    \item On/Off switch
    \item USB-C charging port
    \item 2.54 mm header pins to plug in the AtomS3 (and additional sensor boards), providing power and access to the sensors via an I²C bus
\end{itemize}

With the schematic and Gerber files provided\footnote{https://github.com/teco-kit/HARNode}, the assembled PCB can be ordered online, e.g. at JLCPCB\footnote{\url{https://jlcpcb.com}}. Only the battery needs to be soldered to the PCB; apart from that, no hardware knowledge is required. 

\subsubsection{3D-printed enclosure}

The 3D-printed housing, depicted in \autoref{fig:housing_and_strap}A, fits the PCB and provides, by default, 35 x 55 x 5.5 mm space for a battery, e.g. a Jauch LP523450JU 950 mAh 3.7 V lithium-polymer battery. Like the PCB, the housing is open-source and its STL file is provided, for online production, or by printing yourself\footnote{https://github.com/teco-kit/HARNode}.
The battery is installed by soldering its terminals to the pads on the PCB.
This housing provides a tight fit for the PCB; it slides right in and stays in place, requiring no adhesives.
To the left and right of the housing, there are slots where Velcro straps are attached to mount the HARNode to the body, as explained below.

\subsubsection{Velcro Straps}
\begin{figure}[h]
      \centering
      \includegraphics[width=\linewidth]{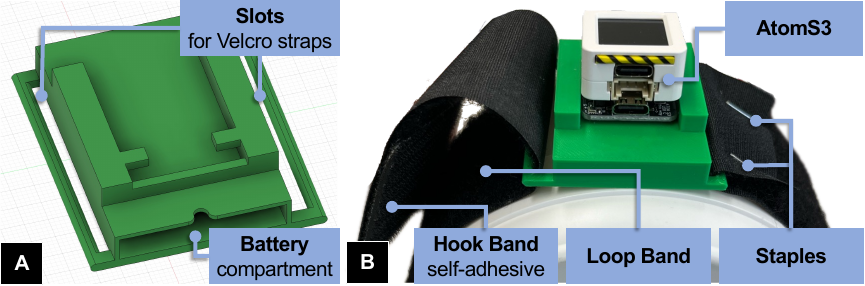}
      \caption{(A) Rendering of the 3D-printed HARNode housing; (B) Assembled HARNode with Velcro straps and labelled components.}
      \label{fig:housing_and_strap}
  \end{figure}

To attach the HARNode to the body, a flexible 50 mm Velcro loop band is used, depicted in \autoref{fig:housing_and_strap}B. The narrower slot in the housing is where the strap is stapled, and the broader side features a pass-through slot for adjusting the strap. After threading the strap through this slot, the Velcro hook strip at the end can be secured to the loop band. For this, 5 cm of 50 mm self-adhesive Velcro hook band is glued onto the end of the loop band.

\begin{figure*}[ht!]
    \centering
    \begin{tikzpicture}
        \node[anchor=south west, inner sep=0] (image) at (0,0) 
            {\includegraphics[width=\linewidth]{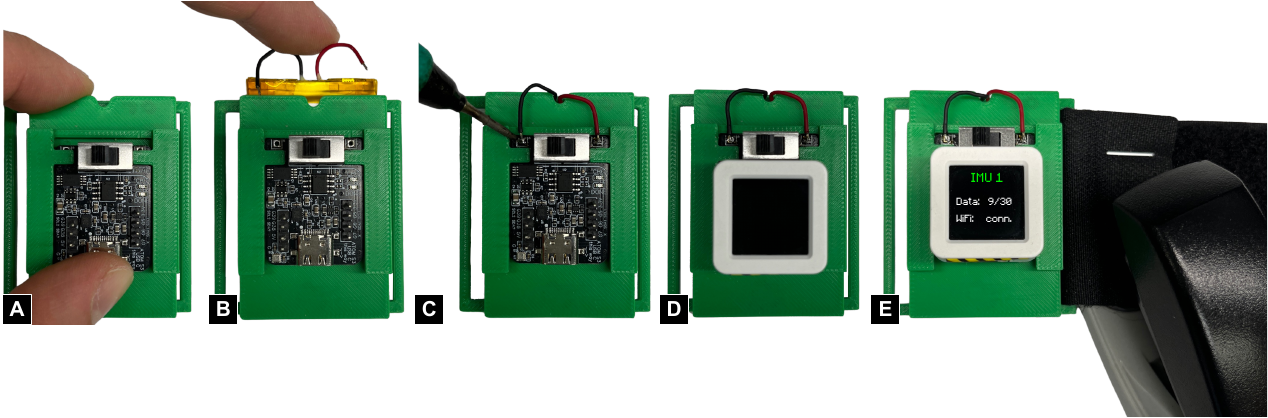}};
        \node[align=justify, anchor=north west, text width=\linewidth] at (0,1.15) {
            \parbox{0.85\linewidth}{
                \captionof{figure}{Assembly process of HARNode: (A) The PCB is plugged into the 3D-printed housing; (B) the battery is inserted into the battery compartment; (C) Both battery terminals are soldered onto the PCB; (D) The AtomS3 is plugged in; (E) The loop band is attached by stapling, and a piece of hook band is glued to its end.}
                \label{fig:assembly}
            }
        };
    \end{tikzpicture}
\end{figure*}

\subsection{Assembly}
The assembly of a HARNode is straightforward and requires only five simple steps. Anyone who is comfortable soldering two wires to the PCB is able to assemble the HARNode. 

\subsubsection{Prerequisites}
To build the HARNode, the items listed in \autoref{tab:required_items} are needed. A stapler and a soldering iron are also required (and a 3D printer if the case is self-printed).
The cost of one HARNode depends on the quantity of required items ordered but does not exceed approximately 48 \euro. The price of 20 assembled PCBs including shipping was 217.95 \euro{} at JLCPCB, which is less than 11~\euro{} per PCB. 

\begin{table}[h]
    \centering
    \caption{Items needed to build a HARNode, including approximate prices.}
        
    \resizebox{\columnwidth}{!}{%
        \begin{tabular}{|c|c|}
\hline
\textbf{Item} & \textbf{Approx. Price (€)} \\
\hline
Assembled PCB & 11 \\
3D printed Housing & <1 \\
Suitable Battery & 10 \\
50 mm Loop Band (per meter) & 3 \\
50 mm Adhesive Hook Band (per meter) & 3 \\
M5Stack Atom3 & 20 \\
\hline
\textbf{Total} & \textbf{48} \\
\hline
            \end{tabular}%
    }

    \label{tab:required_items}
\end{table}

\subsubsection{Assembly Process}
The assembly process is depicted in \autoref{fig:assembly} and consists of five steps: 
1) First, the PCB is plugged into the 3D printed housing by pushing it through the mounting brackets of the housing. The housing ensures a tight fit; no adhesives are required.
2) Then, the battery is inserted into the battery compartment underneath, and 3) both terminals are soldered onto the corresponding pads on the PCB. 
4) Now the AtomS3 can be plugged in, onto the PCB. 
5) Then, the Velcro strap loop band is attached by inserting it into the narrower slot of the housing, using a stapler to secure it. 
Finally, about 5 cm of self-adhesive hook band is glued onto the end of the loop band.

\subsection{Software}
The software needed to operate HARNodes includes the firmware running on the AtomS3's as well as the server that receives the UDP packets sent by the HARNodes. Both have been custom-developed and are released as open-source software\footnote{https://github.com/teco-kit/HARNode}.

\subsubsection{Firmware}
The firmware of the nodes is fully customisable, to tweak the sampling rate and sensor configuration that are used. It is written in C++ using the Arduino framework and is deployed via PlatformIO\footnote{\url{https://platformio.org}}, making it easy to adapt to your needs. Each node connects through a list of predefined Wi-Fi SSIDs that must be provided in the firmware beforehand. In this way, we reduce the setup time while maintaining flexibility to run parallel studies. The node first sends three NTP-requests towards the local server (and interpolates the received times to accommodate for round trip time), to synchronise each node to about 1 ms in average (5 ms max.) \cite{mills2002internet} for standard field setups with one access point and multiple HARNodes. NTP requests are repeated every minute to keep the deviation of the RTC clock in the microsecond level. Once the NTP response is received and the node is synchronised, it starts sending UDP packets with a default of 30 sensor readings per packet, with an adjustable default sampling rate of 166.67 Hz (read every 6 milliseconds), which we found to be a good compromise between motion detection, battery life, and achievable time synchronisation while preventing UDP packet fragmentation \cite{rfc8085}.

\subsubsection{Server}
The NodeJS server provides the NTP responses that each HARNode requests, receives, and stores all UDP data packets, and provides a web-based GUI to start recordings and synchronise a video feed to the HARNodes, provided by the webcam or a connected phone.

\begin{figure}[h]
\includegraphics[width=0.6\linewidth]{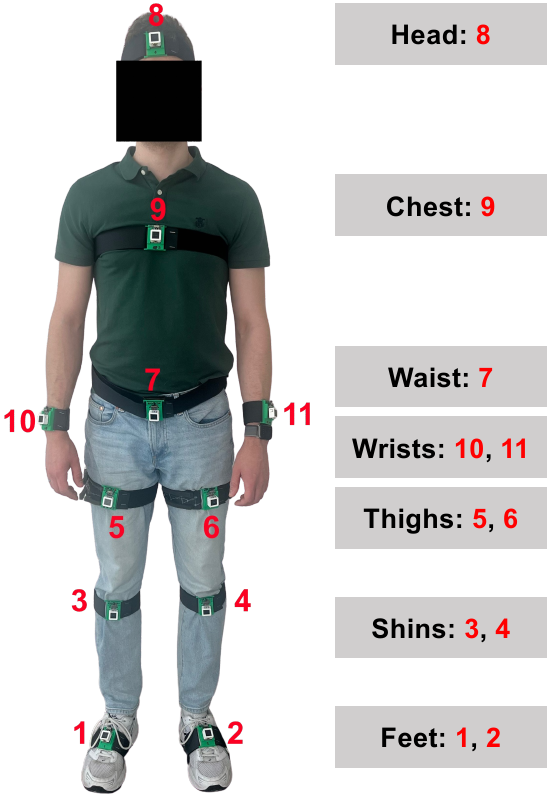}
\caption{Sensor node placement on participant. For our validation study, we used 11 HARNodes, positioned on the feet (1, 2), shins (3, 4), thighs (5, 6), waist (7), head (8), chest (9), and both wrists (10, 11).}
\label{fig:sensor_placement}
\end{figure}


\subsection{Operation}
To use HARNodes, a Wi-Fi access point is required. Any access point can be used, including smartphone-based hotspots or other mobile access points, enabling ad hoc setups outside the laboratory.
The amount of HARNodes usable is limited only by the router's and server's hardware used. For example, phone hotspots are often limited to a few connected devices \cite{apple2024hotspot}, but ordinary Wi-Fi access points can typically handle many more connections.
In our setup, each HARNode sent a packet with 1082 bytes of data at 5.6 Hz (166.67 Hz sampling rate, 30 datapoints of 9-axis 32-bit float IMU data + 8 bit for ID + 8 bit for the number of data points in one packet), which is approximately 6.06 kB/s or $\approx$ 48.48 kbit/s. Even with very conservative estimates of only 5 Mbit/s effective data rate for a 2.4 GHz 802.11 n network (in theory, the maximum possible minimal data rate is 72.2 Mbit/s, up to 600 Mbit/s \cite{5307322}) would easily enable more than 100 nodes: $\frac{5\ \text{Mbit/s}}{48.48\ \text{kbit/s}} \approx 103\ \text{nodes}$.

The Wi-Fi network credentials must be entered in the firmware before flashing. After turning on HARNode, it automatically connects to the entered Wi-Fi network and starts streaming.
The server must be connected to the same Wi-Fi network to receive the sensor streams.

\section{Field Setup and Deployment}
\label{sec:field}
A key feature of our system is the standardised deployment procedure that enables a fast, error-minimised setup in field environments. Before a study session, each sensor node is charged and flashed with the appropriate firmware. The experimenter begins by powering on all HARNodes; each device will show a default identifier on its display. The following steps are then performed to instrument the participant:
\begin{enumerate}
    \item \textbf{Assigning position via the display:} Taking one node at a time, the researcher first scrolls through the numbers of body positions using the built-in button (see \autoref{fig:sensor_placement}). The chosen label appears on the screen, confirming that this node is intended for that location. Then we cycle to the specific position on the body (front, back, left, right (seen from the front)), which currently makes a total of 44 possible positions.
    \item \textbf{Attach node to participant:} The node is placed at the corresponding body location and secured by tightening its Velcro strap. Thanks to the one-size-fits-most strap design, this process only takes a few seconds per sensor and does not require precise alignment or calibration fixtures - the goal is to place the IMU roughly over the intended body segment.
    The display helps to ensure correct orientation: as long as the text on the screen is the right way up, the sensor node is correctly placed.
    \item \textbf{Repeat for all sensors:} Steps 1 - 2 are repeated for each sensor node until the participant is wearing the full set of required IMUs (in our example case: 11 nodes, covering major limbs and torso). The integrated displays help to quickly verify that every intended body location has a sensor and each sensor is correctly identified. It also shows Wi-Fi and server connectivity.
    \item \textbf{Initialise recording:} Once all nodes are in place, the researcher uses the server’s web interface to start the synchronised recording. At this moment, a broadcast command is used to signal the nodes to note the start of an official trial. All sensor data streams are timestamped against the NTP-synchronised clocks as they arrive at the server. The GUI can also initiate a simultaneous video recording if needed. The participant can then perform the experimental tasks while the system logs all the data in the background.
\end{enumerate}
Using this procedure, the entire setup (from donning sensors to beginning data capture) can be completed in just a few minutes, even with a large number of sensors. The protocol is highly repeatable and does not require special technical skills, which facilitates consistent deployment across different operators or study sites. By minimising setup time and complexity, our system lowers the barrier to conduct multi-sensor field studies in realistic environments, enabling researchers to gather rich motion data with ease and reliability.

\section{Validation Study}
To validate the capabilities of the HARNodes and to deliver a contribution to systematically finding relevant positions for IMUs in exoskeletons, we conducted a study to train a machine learning model to precisely detect the moment when somebody walks towards stairs, before their feet even touch the first step.

\subsection{Setup}
Ten participants (mean weight 74.55 kg $\pm$22.46; mean height 1.75 m $\pm$0.10; 10 right-footed, 1 left-footed) each wore 11 HARNodes and completed 10 repetitions of walking towards stair ascent, 10 towards descent, and 2 minutes of level walking. 
The distance from start to the stairs was 2 metres.
HARNodes were strategically positioned on key body locations to capture comprehensive movement data (on both feet, shins, thighs and arms, as well as on the waist, chest and head), as shown in \ref{fig:sensor_placement}.
IMU data were sampled at 166.67 Hz and streamed via Wi-Fi with 30 data points per UDP packet, per HARNode, towards a laptop (MacBook Pro M3) which ran the server software. Each stair-walking activity was recorded on video for manual labelling.

\subsection{Pre-Processing}
The raw sensor data of all 11 HARNodes were aligned and interpolated using the precise, NTP-based, timestamps of each node, and then normalised across all sensors and participants.
Each session was labelled either \textit{"walking"} or \textit{"towards stairs"}, based on the video recordings as ground truth. We define the start of stair-approach intervals as the moment when both legs and feet are aligned in a parallel position, immediately prior to the leading leg initiating the final step before stair ascent; the end of the stair-approach interval as the moment immediately preceding foot placement on the first step in stair ascent, or just before the leading foot moves below the stance foot to contact the step in stair descent.

The continuous data streams were then segmented into fixed-length windows of 25 samples with 75\% overlap to facilitate feature extraction. From these segmented windows, the mean, standard deviation, minimum, maximum, range, median, kurtosis and skew were extracted as features for each sensor and axis (ax, ay, az, gx, gy, gz, mx, my, mz; a: accelerometer, m: magnetometer, g: gyroscope). Left- or right-footedness was also included as a feature for each participant. Balancing of class distribution in the dataset was always performed before model training, by randomly sampling from the under-represent class. This allows us to compare accuracy as a metric, because it is not skewed.

\subsection{Classification}
For the classification task, a random forest machine learning model was used, applying a 70/30 train/test split. Random forests deliver good performance even for our limited amount of training data, which would make it unfeasible to use state-of-the-art deep-learning models such as CNNs and Transformers \cite{alzubaidi2023survey}. 

\section{Evaluation}
\begin{figure}[h]
  \centering
  \includegraphics[width=1\linewidth]{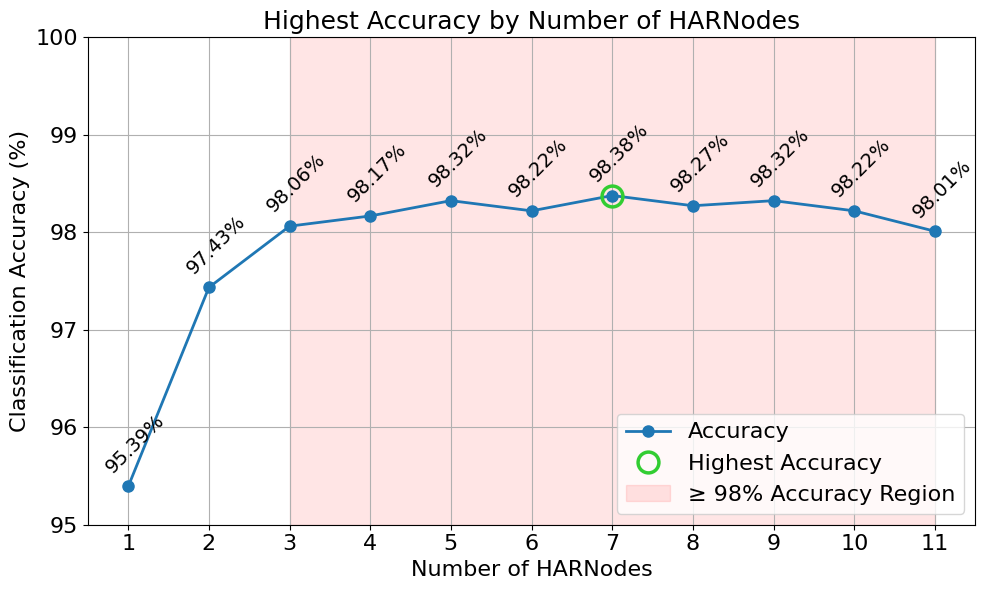}
  \caption{Plot showing number of HARNodes for the classification task vs highest classification accuracy of any combination of n sensors. Using 3-11 nodes results in nearly the same accuracy of $\approx$ 98\%.}
  \label{fig:accuracy_vs_n_sensors}
\end{figure}

To determine both the number and the placement of sensor nodes required for good classification performance, we employed sensor overprovisioning and trained random forest models for binary classification (\textit{"walking"} and \textit{"towards stairs"}) across all possible combinations of sensor positions ($2^{11} = 2048$ models).
For every amount of sensor (1-11), the model with the highest accuracy was calculated, depicted in \autoref{fig:accuracy_vs_n_sensors}.

The best model with all 11 sensor nodes achieves 98.01\% accuracy (Leave-one-subject-out cross-validation mean accuracy: 90.99\% $\pm$ 5.62), while the highest overall accuracy of 98.38\% (LOOCV mean accuracy: 90.53\% $\pm$ 3.74) is achieved with only 7 nodes. Even only using three sensor nodes (\textit{left foot}, \textit{left wrist} and \textit{waist}), results in 98.06\% accuracy (LOOCV mean accuracy: 88.74\% $\pm$ 5.62), surpassing 11 sensor nodes. The detailed precision and recall values for all classifications can be seen in \autoref{tab:detailed_classification}.

\begin{table}
\centering
\caption{Classification accuracy, as well as precision and recall for the binary classification \textit{walking} vs \textit{walking towards stairs}; Acc. = Accuracy; Prec. = Precision; Rec. = Recall.}
\label{tab:detailed_classification}
\begin{tabular}{|c|c|c|c|c|c|}
\hline
\textbf{\# Sensors} & \textbf{Acc. (\%)} & \multicolumn{2}{c|}{\textbf{\textit{Walking} (\%)}} & \multicolumn{2}{c|}{\textbf{\textit{Towards Stairs} (\%)}} \\
\cline{3-6}
 & & \textbf{Prec.} & \textbf{Rec.} & \textbf{Prec.} & \textbf{Rec.} \\
\hline
1 & 95.39 & 98.33 & 92.36 & 92.79 & 98.43 \\
2 & 97.43 & 98.81 & 96.02 & 96.13 & 98.85 \\
3 & 98.06 & 99.25 & 96.86 & 96.93 & 99.27 \\
4 & 98.17 & 99.36 & 96.96 & 97.03 & 99.37 \\
5 & 98.32 & 99.46 & 97.17 & 97.24 & 99.48 \\
6 & 98.22 & 99.36 & 97.07 & 97.13 & 99.37 \\
7 & 98.38 & 99.46 & 97.28 & 97.34 & 99.48 \\
8 & 98.27 & 99.57 & 96.96 & 97.04 & 99.58 \\
9 & 98.32 & 99.57 & 97.07 & 97.14 & 99.58 \\
10 & 98.22 & 99.68 & 96.75 & 96.85 & 99.69 \\
11 & 98.01 & 99.67 & 96.34 & 96.45 & 99.69 \\
\hline
\end{tabular}
\end{table}

\begin{figure}[h]
  \centering
  \includegraphics[width=0.7\linewidth]{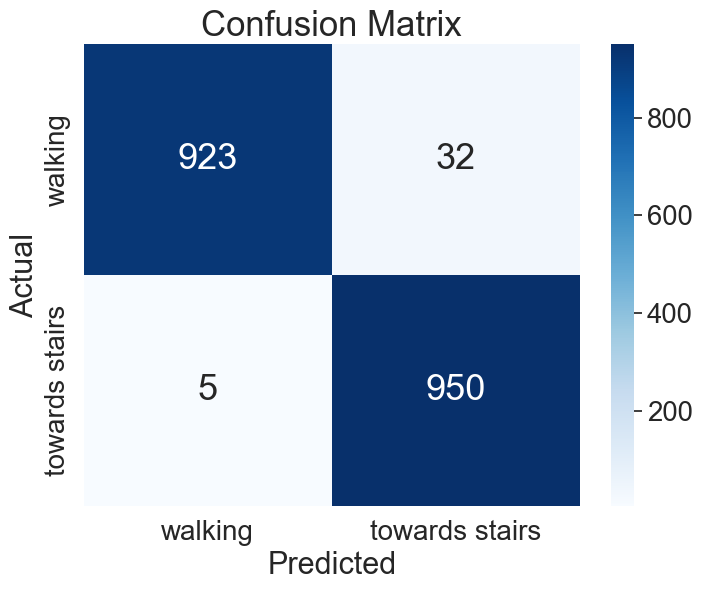}
  \caption{Confusion matrix for binary classification (\textit{walking} vs \textit{towards stairs}) of the model with the highest classification accuracy (7 HARNodes).}
  \label{fig:2f_confusion}
\end{figure}

This clearly shows that only using a small subset of all sensors achieves comparable performance, proving the importance of screening for overprovisioning of sensors when building a human activity recognition system.

That is especially useful in the scenario when you are trying to find the best location for IMUs, for example in exoskeletons. Not only is the performance important, but also the specific location and amount of IMU's required.
We therefore also looked at five specific body positions with practical interest, with a maximum of 4 sensor nodes and computed their classification accuracies for the stair detection task, shown in \autoref{tab:positions}.
For example, a sensor node on the wrist is similar to a watch and socially acceptable; feet, legs, and waist are already covered by the exoskeleton, and a sensor on the head could be hidden inside a hat, for instance. 

\begin{table}[h!]
\centering
\caption{Five HARNode locations of interest, due to their practicality and their respective classification accuracies.}
\begin{tabular}{|l|c|}
\hline
\textbf{Sensor Locations} & \textbf{Accuracy (\%)} \\
\hline
\textit{right foot}, \textit{right thigh}, \textit{right shin}, \textit{waist} & 97.07 \\
\textit{right foot}, \textit{right thigh}, \textit{right wrist} & 96.34 \\
\textit{right foot}, \textit{right thigh} & 95.39 \\
\textit{right foot}, \textit{right wrist}, \textit{head} & 96.49 \\
\textit{right foot} & 90.26 \\
\hline
\end{tabular}
\label{tab:positions}
\end{table}

The results show that with decreasing numbers of sensor nodes on the body, the accuracy drops, but not significantly. Only when relying on only the foot as single sensor placement, we see a sharper decline in accuracy.


\section{Conclusion}
The HARNode system addresses key challenges in the collection of field data for human activity recognition (HAR), in particular the need for high node synchronisation (approx. 1 ms) and the easy handling of a large number of sensors. Based on open-source hardware (ESP32-S3, Bosch BMX160 IMU), it enables quick setup in the field and streaming of data via Wi-Fi. A proof-of-concept study with 10 participants and 11 HARNodes validated the system for recognising the transition to stair climbing and demonstrated the practical value of rapid sensor setup and sensor overprovisioning (11 sensors = 3 selected sensors, $\approx$ 98\% classification accuracy).

Future endeavours include developing a revised body attachment model with more attachment points, automating the calculation of optimal sensor combinations, providing more diverse attachment options beyond Velcro straps, further minimising setup time, and integrating online annotation features during field data collection. All resources about the HARNode system are publicly available under the MIT licence on \url{https://github.com/teco-kit/HARNode}.

\begin{acks}
This work was funded by the Carl Zeiss Foundation through the JuBot project.
\end{acks}

\balance
\bibliographystyle{ACM-Reference-Format}
\bibliography{references}


\begin{thebibliography}{29}


\ifx \showCODEN    \undefined \def \showCODEN     #1{\unskip}     \fi
\ifx \showISBNx    \undefined \def \showISBNx     #1{\unskip}     \fi
\ifx \showISBNxiii \undefined \def \showISBNxiii  #1{\unskip}     \fi
\ifx \showISSN     \undefined \def \showISSN      #1{\unskip}     \fi
\ifx \showLCCN     \undefined \def \showLCCN      #1{\unskip}     \fi
\ifx \shownote     \undefined \def \shownote      #1{#1}          \fi
\ifx \showarticletitle \undefined \def \showarticletitle #1{#1}   \fi
\ifx \showURL      \undefined \def \showURL       {\relax}        \fi
\providecommand\bibfield[2]{#2}
\providecommand\bibinfo[2]{#2}
\providecommand\natexlab[1]{#1}
\providecommand\showeprint[2][]{arXiv:#2}

\bibitem[530(2009)]%
        {5307322}
 \bibinfo{year}{2009}\natexlab{}.
\newblock \showarticletitle{IEEE Standard for Information technology-- Local and metropolitan area networks-- Specific requirements-- Part 11: Wireless LAN Medium Access Control (MAC)and Physical Layer (PHY) Specifications Amendment 5: Enhancements for Higher Throughput}.
\newblock \bibinfo{journal}{\emph{IEEE Std 802.11n-2009 (Amendment to IEEE Std 802.11-2007 as amended by IEEE Std 802.11k-2008, IEEE Std 802.11r-2008, IEEE Std 802.11y-2008, and IEEE Std 802.11w-2009)}} (\bibinfo{year}{2009}), \bibinfo{pages}{1--565}.
\newblock
\href{https://doi.org/10.1109/IEEESTD.2009.5307322}{doi:\nolinkurl{10.1109/IEEESTD.2009.5307322}}


\bibitem[mov(2025)]%
        {movisens}
 \bibinfo{year}{2025}\natexlab{}.
\newblock \bibinfo{title}{{Movisens GmbH}}.
\newblock \bibinfo{howpublished}{\url{https://www.movisens.com/}}.
\newblock
\newblock
\shownote{Accessed: May 21, 2025}.


\bibitem[xse(2025)]%
        {xsens_awinda}
 \bibinfo{year}{2025}\natexlab{}.
\newblock \bibinfo{title}{{MVN Awinda}}.
\newblock \bibinfo{howpublished}{\url{https://www.movella.com/products/motion-capture/xsens-mvn-awinda}}.
\newblock
\newblock
\shownote{Accessed: May 21, 2025}.


\bibitem[nor(2025)]%
        {noraxon}
 \bibinfo{year}{2025}\natexlab{}.
\newblock \bibinfo{title}{{Noraxon: Innovative Biomechanics Tools for Research \& Clinical Use}}.
\newblock \bibinfo{howpublished}{\url{https://www.noraxon.com/}}.
\newblock
\newblock
\shownote{Accessed: May 21, 2025}.


\bibitem[apd(2025)]%
        {apdm_opal}
 \bibinfo{year}{2025}\natexlab{}.
\newblock \bibinfo{title}{{The Opal: a research-grade wearable sensor built for ultimate control}}.
\newblock \bibinfo{howpublished}{\url{https://biofittechs.com/apdm-wearable}}.
\newblock
\newblock
\shownote{Accessed: May 21, 2025}.


\bibitem[ope(2025)]%
        {opensensert_opensim}
 \bibinfo{year}{2025}\natexlab{}.
\newblock \bibinfo{title}{{Wearable and Real-time Kinematics Estimates with OpenSense}}.
\newblock \bibinfo{howpublished}{\url{https://opensimconfluence.atlassian.net/wiki/spaces/OpenSim/pages/53084280/Wearable+and+Real-time+Kinematics+Estimates+with+OpenSense}}.
\newblock
\newblock
\shownote{Accessed: May 21, 2025}.


\bibitem[Alzubaidi et~al\mbox{.}(2023)]%
        {alzubaidi2023survey}
\bibfield{author}{\bibinfo{person}{Laith Alzubaidi}, \bibinfo{person}{Jinshuai Bai}, \bibinfo{person}{Aiman Al-Sabaawi}, \bibinfo{person}{Jose Santamar{\'\i}a}, \bibinfo{person}{Ahmed~Shihab Albahri}, \bibinfo{person}{Bashar Sami~Nayyef Al-Dabbagh}, \bibinfo{person}{Mohammed~A Fadhel}, \bibinfo{person}{Mohamed Manoufali}, \bibinfo{person}{Jinglan Zhang}, \bibinfo{person}{Ali~H Al-Timemy}, {et~al\mbox{.}}} \bibinfo{year}{2023}\natexlab{}.
\newblock \showarticletitle{A survey on deep learning tools dealing with data scarcity: definitions, challenges, solutions, tips, and applications}.
\newblock \bibinfo{journal}{\emph{Journal of Big Data}} \bibinfo{volume}{10}, \bibinfo{number}{1} (\bibinfo{year}{2023}), \bibinfo{pages}{46}.
\newblock


\bibitem[{Apple Inc.}(2024)]%
        {apple2024hotspot}
\bibfield{author}{\bibinfo{person}{{Apple Inc.}}} \bibinfo{year}{2024}\natexlab{}.
\newblock \bibinfo{booktitle}{\emph{How to set up a Personal Hotspot on your iPhone or iPad}}.
\newblock
\urldef\tempurl%
\url{https://support.apple.com/en-us/111785}
\showURL{%
\tempurl}
\newblock
\shownote{Accessed: 2025-05-21}.


\bibitem[Biagetti et~al\mbox{.}(2025)]%
        {Biagetti2025}
\bibfield{author}{\bibinfo{person}{Giorgio Biagetti}, \bibinfo{person}{Michele Sulis}, \bibinfo{person}{Laura Falaschetti}, {and} \bibinfo{person}{Paolo Crippa}.} \bibinfo{year}{2025}\natexlab{}.
\newblock \showarticletitle{High‐Accuracy Clock Synchronization in Low‐Power Wireless sEMG Sensors}.
\newblock \bibinfo{journal}{\emph{Sensors}} \bibinfo{volume}{25}, \bibinfo{number}{3} (\bibinfo{year}{2025}), \bibinfo{pages}{756}.
\newblock
\href{https://doi.org/10.3390/s25030756}{doi:\nolinkurl{10.3390/s25030756}}


\bibitem[Contoli et~al\mbox{.}(2024)]%
        {contoli2024energy}
\bibfield{author}{\bibinfo{person}{Chiara Contoli}, \bibinfo{person}{Valerio Freschi}, {and} \bibinfo{person}{Emanuele Lattanzi}.} \bibinfo{year}{2024}\natexlab{}.
\newblock \showarticletitle{Energy-aware human activity recognition for wearable devices: A comprehensive review}.
\newblock \bibinfo{journal}{\emph{Pervasive and Mobile Computing}} (\bibinfo{year}{2024}), \bibinfo{pages}{101976}.
\newblock


\bibitem[Davoudi et~al\mbox{.}(2021)]%
        {Davoudi2021}
\bibfield{author}{\bibinfo{person}{Anis Davoudi}, \bibinfo{person}{Mamoun~T.\ Mardini}, \bibinfo{person}{Parisa Rashidi}, {and} \bibinfo{person}{Todd Manini}.} \bibinfo{year}{2021}\natexlab{}.
\newblock \showarticletitle{The Effect of Sensor Placement and Number on Physical Activity Recognition and Energy Expenditure Estimation in Older Adults}.
\newblock \bibinfo{journal}{\emph{JMIR mHealth and uHealth}} \bibinfo{volume}{9}, \bibinfo{number}{5} (\bibinfo{year}{2021}), \bibinfo{pages}{e23681}.
\newblock
\href{https://doi.org/10.2196/23681}{doi:\nolinkurl{10.2196/23681}}


\bibitem[Eggert et~al\mbox{.}(2017)]%
        {rfc8085}
\bibfield{author}{\bibinfo{person}{Lars Eggert}, \bibinfo{person}{Godred Fairhurst}, {and} \bibinfo{person}{Greg Shepherd}.} \bibinfo{year}{2017}\natexlab{}.
\newblock \bibinfo{title}{{UDP Usage Guidelines}}.
\newblock \bibinfo{howpublished}{\url{https://datatracker.ietf.org/doc/html/rfc8085}}.
\newblock
\newblock
\shownote{RFC 8085}.


\bibitem[{Espressif Systems}(2021)]%
        {espressif2021esp32s3}
\bibfield{author}{\bibinfo{person}{{Espressif Systems}}.} \bibinfo{year}{2021}\natexlab{}.
\newblock \bibinfo{booktitle}{\emph{ESP32-S3 Series Datasheet}}.
\newblock
\urldef\tempurl%
\url{https://www.espressif.com/sites/default/files/documentation/esp32-s3_datasheet_en.pdf}
\showURL{%
\tempurl}
\newblock
\shownote{Accessed: 2025-05-26}.


\bibitem[Gil-Mart{\'\i}n et~al\mbox{.}(2023)]%
        {gil2023reducing}
\bibfield{author}{\bibinfo{person}{Manuel Gil-Mart{\'\i}n}, \bibinfo{person}{Javier L{\'o}pez-Iniesta}, \bibinfo{person}{Fernando Fern{\'a}ndez-Mart{\'\i}nez}, {and} \bibinfo{person}{Rub{\'e}n San-Segundo}.} \bibinfo{year}{2023}\natexlab{}.
\newblock \showarticletitle{Reducing the impact of sensor orientation variability in human activity recognition using a consistent reference system}.
\newblock \bibinfo{journal}{\emph{Sensors}} \bibinfo{volume}{23}, \bibinfo{number}{13} (\bibinfo{year}{2023}), \bibinfo{pages}{5845}.
\newblock


\bibitem[Hong et~al\mbox{.}(2024)]%
        {hong2024crosshar}
\bibfield{author}{\bibinfo{person}{Zhiqing Hong}, \bibinfo{person}{Zelong Li}, \bibinfo{person}{Shuxin Zhong}, \bibinfo{person}{Wenjun Lyu}, \bibinfo{person}{Haotian Wang}, \bibinfo{person}{Yi Ding}, \bibinfo{person}{Tian He}, {and} \bibinfo{person}{Desheng Zhang}.} \bibinfo{year}{2024}\natexlab{}.
\newblock \showarticletitle{Crosshar: Generalizing cross-dataset human activity recognition via hierarchical self-supervised pretraining}.
\newblock \bibinfo{journal}{\emph{Proceedings of the ACM on Interactive, Mobile, Wearable and Ubiquitous Technologies}} \bibinfo{volume}{8}, \bibinfo{number}{2} (\bibinfo{year}{2024}), \bibinfo{pages}{1--26}.
\newblock


\bibitem[Hussain et~al\mbox{.}(2019)]%
        {hussain2019different}
\bibfield{author}{\bibinfo{person}{Zawar Hussain}, \bibinfo{person}{Michael Sheng}, {and} \bibinfo{person}{Wei~Emma Zhang}.} \bibinfo{year}{2019}\natexlab{}.
\newblock \showarticletitle{Different approaches for human activity recognition: A survey}.
\newblock \bibinfo{journal}{\emph{arXiv preprint arXiv:1906.05074}} (\bibinfo{year}{2019}).
\newblock


\bibitem[Kapsalyamov et~al\mbox{.}(2019)]%
        {kapsalyamov2019state}
\bibfield{author}{\bibinfo{person}{Akim Kapsalyamov}, \bibinfo{person}{Prashant~K Jamwal}, \bibinfo{person}{Shahid Hussain}, {and} \bibinfo{person}{Mergen~H Ghayesh}.} \bibinfo{year}{2019}\natexlab{}.
\newblock \showarticletitle{State of the art lower limb robotic exoskeletons for elderly assistance}.
\newblock \bibinfo{journal}{\emph{IEEE Access}}  \bibinfo{volume}{7} (\bibinfo{year}{2019}), \bibinfo{pages}{95075--95086}.
\newblock


\bibitem[Khan et~al\mbox{.}(2022)]%
        {Li2022}
\bibfield{author}{\bibinfo{person}{Ahsan~Raza Khan}, \bibinfo{person}{Habib~Ullah Manzoor}, {and} \bibinfo{person}{Ahmed Zoha}.} \bibinfo{year}{2022}\natexlab{}.
\newblock \showarticletitle{A Privacy and Energy‐Aware Federated Framework for Human Activity Recognition}.
\newblock \bibinfo{journal}{\emph{Sensors}} \bibinfo{volume}{23}, \bibinfo{number}{23} (\bibinfo{year}{2022}), \bibinfo{pages}{9339}.
\newblock
\href{https://doi.org/10.3390/s23239339}{doi:\nolinkurl{10.3390/s23239339}}


\bibitem[Kunze and Lukowicz(2014)]%
        {kunze2014sensor}
\bibfield{author}{\bibinfo{person}{Kai Kunze} {and} \bibinfo{person}{Paul Lukowicz}.} \bibinfo{year}{2014}\natexlab{}.
\newblock \showarticletitle{Sensor placement variations in wearable activity recognition}.
\newblock \bibinfo{journal}{\emph{IEEE Pervasive Computing}} \bibinfo{volume}{13}, \bibinfo{number}{4} (\bibinfo{year}{2014}), \bibinfo{pages}{32--41}.
\newblock


\bibitem[Mills(2002)]%
        {mills2002internet}
\bibfield{author}{\bibinfo{person}{David~L Mills}.} \bibinfo{year}{2002}\natexlab{}.
\newblock \showarticletitle{Internet time synchronization: the network time protocol}.
\newblock \bibinfo{journal}{\emph{IEEE Transactions on communications}} \bibinfo{volume}{39}, \bibinfo{number}{10} (\bibinfo{year}{2002}), \bibinfo{pages}{1482--1493}.
\newblock


\bibitem[Moreira et~al\mbox{.}(2022)]%
        {moreira2022review}
\bibfield{author}{\bibinfo{person}{Lu{\'\i}s Moreira}, \bibinfo{person}{Joana Figueiredo}, \bibinfo{person}{Jo{\~a}o Cerqueira}, {and} \bibinfo{person}{Cristina~P Santos}.} \bibinfo{year}{2022}\natexlab{}.
\newblock \showarticletitle{A review on locomotion mode recognition and prediction when using active orthoses and exoskeletons}.
\newblock \bibinfo{journal}{\emph{Sensors}} \bibinfo{volume}{22}, \bibinfo{number}{19} (\bibinfo{year}{2022}), \bibinfo{pages}{7109}.
\newblock


\bibitem[Ni et~al\mbox{.}(2024)]%
        {Ni2024}
\bibfield{author}{\bibinfo{person}{Jianyuan Ni}, \bibinfo{person}{Hao Tang}, \bibinfo{person}{Syed~T.\ Haque}, \bibinfo{person}{Yan Yan}, {and} \bibinfo{person}{Anne H.\~H.\ Ngu}.} \bibinfo{year}{2024}\natexlab{}.
\newblock \showarticletitle{A Survey on Multimodal Wearable Sensor‐Based Human Action Recognition}.
\newblock \bibinfo{journal}{\emph{arXiv e‐prints}} (\bibinfo{year}{2024}).
\newblock
\showeprint{2404.15349}


\bibitem[Pesenti et~al\mbox{.}(2023)]%
        {pesenti2023imu}
\bibfield{author}{\bibinfo{person}{Mattia Pesenti}, \bibinfo{person}{Giovanni Invernizzi}, \bibinfo{person}{Julie Mazzella}, \bibinfo{person}{Marco Bocciolone}, \bibinfo{person}{Alessandra Pedrocchi}, {and} \bibinfo{person}{Marta Gandolla}.} \bibinfo{year}{2023}\natexlab{}.
\newblock \showarticletitle{IMU-based human activity recognition and payload classification for low-back exoskeletons}.
\newblock \bibinfo{journal}{\emph{Scientific Reports}} \bibinfo{volume}{13}, \bibinfo{number}{1} (\bibinfo{year}{2023}), \bibinfo{pages}{1184}.
\newblock


\bibitem[Slade et~al\mbox{.}(2021)]%
        {slade2021open}
\bibfield{author}{\bibinfo{person}{Patrick Slade}, \bibinfo{person}{Ayman Habib}, \bibinfo{person}{Jennifer~L Hicks}, {and} \bibinfo{person}{Scott~L Delp}.} \bibinfo{year}{2021}\natexlab{}.
\newblock \showarticletitle{An open-source and wearable system for measuring 3D human motion in real-time}.
\newblock \bibinfo{journal}{\emph{IEEE Transactions on Biomedical Engineering}} \bibinfo{volume}{69}, \bibinfo{number}{2} (\bibinfo{year}{2021}), \bibinfo{pages}{678--688}.
\newblock


\bibitem[Sposito et~al\mbox{.}(2022)]%
        {sposito2022exoskeletons}
\bibfield{author}{\bibinfo{person}{Matteo Sposito}, \bibinfo{person}{Tommaso Poliero}, \bibinfo{person}{Christian Di~Natali}, \bibinfo{person}{Marianna Semprini}, \bibinfo{person}{Giacinto Barresi}, \bibinfo{person}{Matteo Laffranchi}, \bibinfo{person}{Darwin~Gordon Caldwell}, \bibinfo{person}{Lorenzo De~Michieli}, {and} \bibinfo{person}{Jes{\'u}s Ortiz}.} \bibinfo{year}{2022}\natexlab{}.
\newblock \showarticletitle{Exoskeletons in elderly healthcare}.
\newblock In \bibinfo{booktitle}{\emph{Internet of Things for Human-Centered Design: Application to Elderly Healthcare}}. \bibinfo{publisher}{Springer}, \bibinfo{pages}{353--374}.
\newblock


\bibitem[Tsukamoto et~al\mbox{.}(2023)]%
        {tsukamoto2023best}
\bibfield{author}{\bibinfo{person}{Akihisa Tsukamoto}, \bibinfo{person}{Naoto Yoshida}, \bibinfo{person}{Tomoko Yonezawa}, \bibinfo{person}{Kenji Mase}, {and} \bibinfo{person}{Yu Enokibori}.} \bibinfo{year}{2023}\natexlab{}.
\newblock \showarticletitle{Where Are the Best Positions of IMU Sensors for HAR?-Approach by a Garment Device with Fine-Grained Grid IMUs}. In \bibinfo{booktitle}{\emph{Adjunct Proceedings of the 2023 ACM International Joint Conference on Pervasive and Ubiquitous Computing \& the 2023 ACM International Symposium on Wearable Computing}}. \bibinfo{pages}{445--450}.
\newblock


\bibitem[\url{https://www.bioxgroup.dk/products-biox-armband/}({[n.\,d.]})]%
        {biox_bands_researchgate}
\bibfield{author}{\bibinfo{person}{\url{https://www.bioxgroup.dk/products-biox-armband/}}.} \bibinfo{year}{[n.\,d.]}\natexlab{}.
\newblock \bibinfo{title}{{BioXBands}}.
\newblock
\newblock
\shownote{Accessed: May 21, 2025}.


\bibitem[Zeagler(2017)]%
        {zeagler2017wear}
\bibfield{author}{\bibinfo{person}{Clint Zeagler}.} \bibinfo{year}{2017}\natexlab{}.
\newblock \showarticletitle{Where to wear it: functional, technical, and social considerations in on-body location for wearable technology 20 years of designing for wearability}. In \bibinfo{booktitle}{\emph{Proceedings of the 2017 ACM International Symposium on Wearable Computers}}. \bibinfo{pages}{150--157}.
\newblock


\bibitem[Zhang et~al\mbox{.}(2022)]%
        {zhang2022deep}
\bibfield{author}{\bibinfo{person}{Shibo Zhang}, \bibinfo{person}{Yaxuan Li}, \bibinfo{person}{Shen Zhang}, \bibinfo{person}{Farzad Shahabi}, \bibinfo{person}{Stephen Xia}, \bibinfo{person}{Yu Deng}, {and} \bibinfo{person}{Nabil Alshurafa}.} \bibinfo{year}{2022}\natexlab{}.
\newblock \showarticletitle{Deep learning in human activity recognition with wearable sensors: A review on advances}.
\newblock \bibinfo{journal}{\emph{Sensors}} \bibinfo{volume}{22}, \bibinfo{number}{4} (\bibinfo{year}{2022}), \bibinfo{pages}{1476}.
\newblock


\end{thebibliography}

\appendix

\end{document}